\documentclass[%
    a4paper,
    aip, 
    jmp, 
    amsmath,amssymb,
    reprint, onecolumn
]{revtex4-2}

\usepackage[utf8]{inputenc}
\usepackage[T1]{fontenc}
\usepackage{mathptmx}

\usepackage{stmaryrd}                           
\usepackage[cal=boondoxo,scr=pxtx]{mathalfa}    
\usepackage{amsthm}
    \theoremstyle{plain}
        
        \newtheorem{property}{Property}

    \theoremstyle{definition}
        \newtheorem{definition}{Definition}
        \newtheorem*{notation}{Notation}

    \theoremstyle{remark}
        \newtheorem*{remark}{Remark}

\usepackage[colorlinks,hyperindex]{hyperref}  
\hypersetup{
   colorlinks,%
   citecolor=blue,%
   linkcolor=blue,%
   urlcolor=blue,%
}
 
\newcommand{\llbra}{\llbracket}
\newcommand{\rrket}{\rrbracket}
\renewcommand{\leq}{\leqslant}

\newcommand{\sgn}{\mathrm{sgn}}
\newcommand{\Lie}{\ensuremath{\mathscr{L}}}
\newcommand{\ML}{\Theta}
\newcommand{\sss}[1]{\scriptscriptstyle{#1}}

\begin{document}

\title{The Schouten--Nijenhuis bracket, codifferential of products and generalized interior products of {\em p}-forms}

\author{E.~Huguet}
    \affiliation{Universit\'e Paris Cité, APC-Astroparticule et Cosmologie (UMR-CNRS 7164), 
    Batiment Condorcet, 10 rue Alice Domon et L\'eonie Duquet, F-75205 Paris Cedex 13, France.}
    \email{huguet@apc.univ-paris7.fr}
\author{J.~Queva}
    \affiliation{Universit\'e de Corse -- CNRS UMR 6134 SPE, Campus Grimaldi BP 52, 20250 Corte, France.}
    \email{queva@univ-corse.fr}
\author{J.~Renaud}
    \affiliation{APC-Astroparticule et Cosmologie (UMR-CNRS 7164), 
    Batiment Condorcet, 10 rue Alice Domon et L\'eonie Duquet, F-75205 Paris Cedex 13, France.}
    \email{jrenaud@apc.in2p3.fr}

\date{\today}

\begin{abstract}
    Identities pertaining to the de Rham codifferential $\delta$ in differential geometry are scattered in the literature.
    This article gathers such formulas involving usual differential operators (Lie derivative, Schouten--Nijenhuis bracket, etc.), while adding some new ones using a natural extension of the interior product, to provide a compact handy summary.
\end{abstract}

\maketitle

\section{Introduction}

Computations with $p$-forms involve differential operators such as the exterior derivative $d$ and the Lie derivative $\Lie_v$, both fulfilling a Leibniz identity with respect to the exterior product $\wedge$.
In stark contrast is the codifferential $\delta$ which is built out of $d$ and the Hodge operator $*$ and for which there is no such identity.
However, some other identities involving $\delta$ can be found scattered in the literature, often in a wider setting.
This short paper aims to gather such formulas in one compact space making it a useful toolbox for the practitioners of the field.
Those formulas are extended further using a generalized interior product, producing new (up to our knowledge) and useful formulas such as Eqs.~\eqref{eq:[d,jA]B}--\eqref{eq:[iA,box]B}.
Our interest in such formulas came as a by-product of our studies on restrictions of differential operators from flat ambient space to cosmological spaces, more precisely computations needed for the generalization to $p$-forms of earlier work\cite{Huguet:2022rxi}.

This article is organized as follows.
Section~\ref{sec:Schouten} deals first with the Schouten--Nijenhuis (SN) bracket, brought back to forms, and its properties,
then with its relationship with the codifferential $\delta$.
Some explicit and little known formulas are gleaned along the way.
In Sec.~\ref{sec:Produit_Generalise} the interior product $i_v$ is generalized in a natural manner with respect to the Hodge operator. 
This provides new identities and in peculiar it involves naturally the Schouten--Nijenhuis bracket.
Noteworthy and new formulas involving the differential operators $d$, $\Lie_v$, $\delta$ and $\square$ are then derived.
Finally, our conventions regarding differential geometry follow those of Ref.~\onlinecite{Fecko} and needed formulas are briefly recalled in appendix~\ref{app:Overview_Geo_Diff}.
Proofs and the outline of some longer calculations are gathered in appendix~\ref{app:Long_Computations} in order not to clutter the main body of the paper.
\begin{notation}
    In the following paper, the $n$-dimensional oriented pseudo-Riemannian manifold is written as $(M,g)$ with $g$ its metric.
    The usual interior and exterior product $i_v$ and  $j_v$, see App.~\ref{app:Overview_Geo_Diff}, are extended to $1$-forms in a natural way.
    For $\lambda\in\Omega^1$ and $\beta\in\Omega^b$ we note
    \begin{align}
        &i^\lambda\beta = i_{\sharp\lambda}\beta,\\
        &j^\lambda\beta = j_{\sharp\lambda}\beta = \lambda\wedge\beta.
    \end{align}
    They correspond to each other through the Hodge operator
    \begin{equation}\label{eq:springboard1}
    i^\lambda\beta = (-1)^{b+1} *^{-1}j^\lambda*\beta.
\end{equation}
These  products will be further generalized in Sec.~\ref{sec:Produit_Generalise} with definition~\ref{def:Prod_Gene}.
\end{notation}
\begin{notation}
    For ease of use, given an integer $n$, the (anti-)commutator between two operators $X$ and $Y$ is written as
    \begin{equation}
        [X,Y]_n = XY - (-1)^nYX,
    \end{equation}
    recovering the commutator for even values of $n$ and anticommutator for odd values.
    When there is no ambiguity we set, as usual, $[X,Y] = [X,Y]_0 = XY - YX$ and $[X,Y]_+ = [X,Y]_1 = XY + YX$.
\end{notation}

\section{The Schouten--Nijenhuis bracket on forms and its properties}
\label{sec:Schouten}

Schouten\cite{Schouten:1940,Schouten:1953}, and then Schouten--Nijenhuis\cite{Nijenhuis:1955A,Nijenhuis:1955B}, bracket came from the natural endeavor in differential geometry of generalizing the Lie derivative not just along vectors but also along tensors.
In particular, it is the natural extension of the Lie derivative to multivectors\cite{Tulczyjew:1974,Michor1987} (i.e. sections of $\smash{\bigwedge^pTM}$).
In addition to this, it fulfills a wealth of algebraic relations making it one of the simplest incarnation of a Gerstenhaber algebra\cite{Gerstenhaber:1963,Kosmann-Schwarzbach:1995,Roger:2009}.
For a beautiful overview regarding Schouten brackets and its generalizations we refer the reader to Ref.~\onlinecite{kosmannschwarzbach2021schouten}.

In this section, we expose how, once brought back to forms, it arises out of the codifferential of the product of two forms.

\begin{definition}\label{def:Crochet_Schouten}
The Schouten--Nijenhuis bracket\cite{Marle97}, brought back to forms, is the unique bilinear application fulfilling:
\begin{alignat}{2}
    &\llbra\alpha,\beta\rrket 
        = -(-1)^{(a-1)(b-1)}\llbra\beta,\alpha\rrket
        \in \Omega^{a+b-1},
        &&\alpha\in\Omega^a,\ \beta\in\Omega^b,
        \label{eq:SchoutenPermutation}\\
    &\llbra\alpha,\beta\wedge\gamma\rrket
        = \llbra\alpha,\beta\rrket\wedge\gamma
        + (-1)^{(a-1)b}\beta\wedge\llbra\alpha,\gamma\rrket,
        &\quad& \alpha\in\Omega^a,\ \beta\in\Omega^b,\ \gamma\in\Omega^c,
        \label{eq:Schouten_triple}\\
    &\llbra \phi,\psi\rrket = 0,
        &&\phi,\psi\in\Omega^0,\\
    &\llbra\lambda,\beta\rrket
        = \flat(\Lie_{\sharp\lambda} \sharp\beta),
        &&\lambda\in\Omega^1,\ \beta\in\Omega^b.\label{eq: one-form}
\end{alignat}
\end{definition}
The ``usual'' Schouten--Nijenhuis bracket is naturally defined on multivectors, it is lowered to forms using the musical applications $\flat$ and $\sharp$
coming from the metric $g$ (see App.~\ref{app:Overview_Geo_Diff}), namely, $\llbra\alpha,\beta\rrket=\flat [\sharp\alpha,\sharp\beta]_{\sss{SN}}$, where $[\ ,\ ]_{\sss{SN}}$ stands for the usual Schouten--Nijenhuis bracket and for the usual commutator if $\alpha$ and $\beta$ are 1-forms.
For $u$ and $v$ two vectors it reads as $\flat [u,v] = \llbra\flat u, \flat v\rrket$.
Nevertheless, this bracket has some particularities compared to the usual one, due to the Lie derivative acting on the $\sharp$ operator. In fact Eq.~\eqref{eq: one-form} leads to
\begin{equation}\label{eq:L_gtilde}
    \llbra\lambda,\beta\rrket
        = (\Lie_{\sharp\lambda}\tilde g)^{mn}j_m i_n\beta
    +\Lie_{\sharp\lambda}\beta,   
\end{equation}
where $\lambda$ is a $1$-form, $\beta$ is a $b$-form, $\tilde g$ is the corresponding metric on forms and $j_m$, $i_n$ stand for $j_{e_m}$ and $i_{e_n}$ (see App.~\ref{app:Long_Computations}).
When $\sharp\lambda$ is a Killing vector it simplifies to 
$\llbra\lambda,\beta\rrket
        = \Lie_{\sharp\lambda}\beta$.
An explicit formula for the SN bracket on $p$-forms is then given by the following formula (see App.~\ref{app:Long_Computations} for a proof).
\begin{property}\label{prop:Crochet_explicite}
    Let $\alpha\in\Omega^a$ and $\beta\in\Omega^b$ be two forms, and $e^k$ a co-frame, setting $c = a + b -1$, then
    \begin{align}
            \llbra\alpha,\beta\rrket
            = \frac{1}{a!b!}\Bigl\{&
            a \alpha_{k_1\cdots k_{a-1} r} (d\beta_{k_a\cdots k_c}, e^r)
            -b (d\alpha_{k_1\cdots k_{a}}, e^r)\beta_{k_{a+1}\cdots k_c r}\nonumber\\
            &+ ab \alpha_{k_1\cdots k_{a-1}r}\beta_{k_a\cdots k_{c-1} s}
            \tilde c^{rs}_{k_c}
            \Bigr\}e^{k_1}\wedge\cdots\wedge e^{k_c},\label{eq:Crochet_arbitraire_explicite}
    \end{align}
        where $\tilde c^{rs}_k=g^{ri}g^{sj}g_{kl}c_{ij}^l$ and $c$ are the anholonomy coefficients of the frame.
\end{property}
\begin{remark}
    The last term in Eq.~\eqref{eq:Crochet_arbitraire_explicite} vanishes as soon as the frame is holonomic.
\end{remark}

This bracket fulfills a broad set of identities which we recall or derive below.
They are heavily used later on.
\begin{property}
    The Schouten--Nijenhuis bracket fulfills a graded Jacobi identity\cite{Nijenhuis:1955A,Marle97,Mikami2001}
    \begin{equation}\label{eq:Jacobi_gradue}
        (-1)^{(a-1)(c-1)}
        \llbra\alpha,\llbra\beta,\gamma\rrket\rrket +
        (-1)^{(b-1)(a-1)}
        \llbra\beta,\llbra\gamma,\alpha\rrket\rrket +
        (-1)^{(c-1)(b-1)}
        \llbra\gamma,\llbra\alpha,\beta\rrket\rrket = 0.
    \end{equation}
\end{property}

Equations \eqref{eq:SchoutenPermutation}, \eqref{eq:Schouten_triple} and \eqref{eq:Jacobi_gradue},
that are graded antisymmetry, graded Leibniz identity and graded Jacobi identity,
make the bracket a Gerstenhaber algebra.
As usual, the graded Jacobi identity can be written in the suggestive form
\begin{equation}
    \llbra\alpha,\llbra\beta,\gamma\rrket\rrket
    = \llbra\llbra\alpha,\beta\rrket,\gamma\rrket
    + (-1)^{(a-1)(b-1)}\llbra\beta,\llbra\alpha,\gamma\rrket\rrket,
\end{equation}
making $\llbra\alpha,\cdot\rrket$ a derivative of degree $a-1$ with respect to $\Omega$ equipped with the Schouten--Nijenhuis bracket (the ``Schouten degree'' of a form being its exterior degree shifted by minus one).

\begin{remark}
    For $\alpha\in\Omega^a$, $\beta\in\Omega^b$ and $\gamma\in\Omega^c$ through elementary algebra one gets,
    in order to expand the bracket with respect to the left argument, the following useful identity:
    \begin{align}
        \llbra\alpha\wedge\beta,\gamma\rrket
        &= \alpha\wedge\llbra\beta,\gamma\rrket
        + (-1)^{ab}\beta\wedge\llbra\alpha,\gamma\rrket.
        \label{eq:AB-C_gauche}
    \end{align}
    For $\phi\in\Omega^0$ a function and $\lambda\in\Omega^1$ a $1$-form notice that
    \begin{equation}\label{eq:Bracket_Scalaire_1forme}
        \llbra\phi,\lambda\rrket
        = -\llbra\lambda,\phi\rrket
        = - \Lie_{\sharp\lambda}\phi
        = - i^\lambda d\phi
        = - (\lambda, d\phi)
        = - i^{d\phi}\lambda,
    \end{equation}
    which extends by induction on the degree to
    \begin{equation}\label{eq:Bracket_Scalaire_FormeGene}
        \llbra\phi,\alpha\rrket
        = - i^{d\phi}\alpha
        = (-1)^a\llbra\alpha,\phi\rrket,\quad
        \phi\in\Omega^0,\ \alpha\in\Omega^a.
    \end{equation}
\end{remark}
\begin{remark}
    Taking into account the commutations relations between $i^\lambda$ and $j^\lambda$ with the Hodge operator, cf. Eqs.~\eqref{eq:*i_vA} and \eqref{eq:*j_vA}, provides a noteworthy identity.
    First, note that for $\lambda,\tau\in\Omega^1$ two $1$-forms and $\gamma\in\Omega^c$ one has
    \begin{equation}\label{eq:*L*BC} 
        *^{-1}\Lie_{\sharp\tau}*\lambda\wedge\gamma
        = *^{-1}\Lie_{\sharp\tau}*j^\lambda\gamma
        = j^\lambda *^{-1}\Lie_{\sharp\tau}*\gamma
        + \llbra\tau,\lambda\rrket\wedge\gamma,
    \end{equation}
    by commuting $j^\lambda$ with $*$ and then noticing that
    $\Lie_{\sharp\tau}i_{\sharp\lambda}
    = [\Lie_{\sharp\tau},i_{\sharp\lambda}] + i_{\sharp\lambda}\Lie_{\sharp\tau}
    = i_{[\sharp\tau,\sharp\lambda]} + i_{\sharp\lambda}\Lie_{\sharp\tau}
    = i_{\Lie_{\sharp\tau}\sharp\lambda} + i_{\sharp\lambda}\Lie_{\sharp\tau}$,
    then commuting $i^\lambda=i_{\sharp\lambda}$ and $i_{\Lie_{\sharp\tau}\sharp\lambda}$ with $*^{-1}$ while remembering that
    $\flat\Lie_{\sharp\tau}\sharp\lambda
    = \llbra\tau,\lambda\rrket$.
    Now for the very special case in which $\gamma = 1\in\Omega^0$ one has
    \begin{align}
        *^{-1}\Lie_{\sharp\tau}*\lambda 1
        &= \lambda *^{-1}\Lie_{\sharp\tau}*1
        + \llbra\tau,\lambda\rrket
        = \lambda *^{-1}\Lie_{\sharp\tau}\omega
        + \llbra\tau,\lambda\rrket\\
        &= -\lambda *^{-1}(\delta\flat\sharp\tau)\omega
        + \llbra\tau,\lambda\rrket
        = -(\delta\tau)\lambda
        + \llbra\tau,\lambda\rrket.
    \end{align}
    Then pushing the degree from $1$ to an arbitrary degree is achieved by using Eq.~\eqref{eq:*L*BC},
    reinstating a coefficient $\phi\in\Omega^0$ with $\llbra\tau,\phi\beta\rrket = \llbra\tau,\phi\rrket\beta + \phi\llbra\tau,\beta\rrket$, giving rise to
    \begin{equation}\label{eq:*L*B}
        *^{-1}\Lie_{\sharp\tau}*\beta
        = \llbra\tau,\beta\rrket
        - (\delta\tau)\beta,
    \end{equation}
    or simply put
    \begin{equation}
        \llbra\lambda,\beta\rrket
        = \flat\Lie_{\sharp\lambda}\sharp\beta
        = *^{-1}\Lie_{\sharp\lambda}*\beta
        + (\delta\lambda)\beta,\quad
        \lambda\in\Omega^1,\
        \beta\in\Omega^b.
    \end{equation}
    Stripping the $b$-form and setting $v=\sharp\lambda$ this finally yields the identity
    \begin{equation}
        *^{-1}\Lie_{v}*
        = \flat\Lie_{v}\sharp
        + \mathrm{div}(v)\mathrm{Id}.
    \end{equation}
\end{remark}

The Schouten--Nijenhuis bracket is the measure of the failure of $\delta$ to be a derivation with respect to the exterior product $\wedge$, which is embodied in the following formula.  

\begin{property}\label{thm:delta_Schouten} 
    Let $\alpha\in\Omega^a$ and $\beta\in\Omega^b$, then
    \begin{equation}
        \label{eq:delta_Schouten}
        \delta(\alpha\wedge\beta)
        = (\delta\alpha)\wedge\beta
        + (-1)^a\alpha\wedge(\delta\beta)
        + (-1)^a\llbra \alpha, \beta\rrket.
   \end{equation}
\end{property}

Equation~Eq.~\eqref{eq:delta_Schouten} or Eq.~\eqref{eq:Representation_Bracket} below, has been known for a long time\cite{Koszul:1985,Vaisman1990,vaisman2012lectures} and can be recovered from Gerstenhaber original work\cite{Gerstenhaber:1963} through Hochschild-Kostant-Rosenberg theorem\cite{HKR:1962}, or
    be found by other means\cite{Coll:2003ym}.
However in those earlier works this kind of identity is not cast in the ``simple'' framework of $p$-forms and it requires some effort to bring it to Eq.~\eqref{eq:delta_Schouten}.
Precisely, Koszul in Ref.~\onlinecite{Koszul:1985}, following the proof of Lemma~(2.1), explicitly constructs a derivative noted $D_\nabla$ such that a formula equivalent to Eq.~\eqref{eq:Representation_Bracket} holds on multivectors.
Comparing the derivative $D_\nabla$ to $i^m\nabla_m$ in Eq.~\eqref{eq:Koszul} shows that both are related through the musical operators $\sharp$ and $\flat$.
Vaisman in Refs.~\onlinecite{Vaisman1990} and \onlinecite{vaisman2012lectures}, following Koszul \cite{Koszul:1985}, derives Eq.~\eqref{eq:delta_Schouten} by noticing that $D_\nabla = -\sharp\delta\flat$ (see Eq.~(1.35) directly on forms of Ref.~\onlinecite{vaisman2012lectures}).
Note that Ref.~\onlinecite{vaisman2012lectures} being a textbook is fully detailed, while in Ref.~\onlinecite{Vaisman1990} a careful reading of the paragraph between Eqs.~(1.6) and (1.7) is needed to obtain the formula.
Coll and Ferrando in Ref.~\onlinecite{Coll:2003ym} obtain on multivectors a relation equivalent to Eq.~\eqref{eq:delta_Schouten}, Eq.~(13) there, using the properties of the Leibniz bracket $\{\ ,\ \}_{\delta}$ (see Ref.~\onlinecite{Coll:2003ym}) on a graded algebra.
Finally, in order to be self-contained, we provide in App.~\ref{app:Long_Computations} a direct proof of Eq.~\eqref{eq:delta_Schouten} relying on Eqs.~\eqref{eq:Bracket_Scalaire_FormeGene} and \eqref{eq:*L*B} with an induction on the degree of the form.
\begin{remark}
    Equation~\eqref{eq:delta_Schouten} provides an alternative representation of the bracket as it generates it
    \begin{equation}\label{eq:Representation_Bracket}
        \llbra\alpha,\beta\rrket
        = (-1)^a\left[
            \delta(\alpha\wedge\beta)
            - (\delta\alpha)\wedge\beta
            - (-1)^a\alpha\wedge(\delta\beta)
        \right],
    \end{equation}
    and makes the Gerstenhaber algebra an \emph{exact} Gerstenhaber algebra, that is a Batalin-Vilkovisky algebra\cite{Batalin:1981,Witten:1990}.
    As the bracket is generated by $\delta$ it behaves nicely with respect to it, namely,
    \begin{equation}\label{eq:DeltaBracket}
        \delta\llbra\alpha,\beta\rrket
        = \llbra\delta\alpha,\beta\rrket
        - (-1)^a\llbra\alpha,\delta\beta\rrket.
    \end{equation}
    In addition making use of Eqs.~\eqref{eq:Representation_Bracket} and \eqref{eq:DeltaBracket},  of the graded antisymmetry and Leibniz identity
    allows for a straightforward proof of the graded Jacobi identity Eq.~\eqref{eq:Jacobi_gradue}.
    Moreover Eq.~\eqref{eq:Representation_Bracket} gives rise to another representation of the bracket\cite{Nijenhuis:1955B}:
    \begin{equation}
        \llbra\alpha,\beta\rrket
        =-(\nabla_m\alpha)\wedge i^m\beta
        + (-1)^{a+1}(i^m\alpha)\wedge\nabla_m\beta,
    \end{equation}
    where $\nabla$ stands for the Levi--Civita connection.
    Indeed we have
    \begin{align}
         \llbra\alpha,\beta\rrket
       & = (-1)^a\left[
            -i^m\nabla_m(\alpha\wedge\beta)
            - (-i^m\nabla_m\alpha)\wedge\beta
            -(-1)^a\alpha\wedge(-i^m\nabla_m\beta)
            \right]\label{eq:Koszul}\\ 
        &= (-1)^a[-(i^m\nabla_m\alpha)\wedge\beta
            -(-1)^a (\nabla_m\alpha)\wedge i^m\beta
            -(i^m\alpha)\wedge(\nabla_m\beta)\nonumber\\
        &\qquad\qquad\quad
            -(-1)^a\alpha\wedge (i^m\nabla_m\beta)
            +(i^m\nabla_m\alpha)\wedge\beta
            +(-1)^a\alpha\wedge(i^m\nabla_m\beta) ]\\
        &=-(\nabla_m\alpha)\wedge i^m\beta
        + (-1)^{a+1}(i^m\alpha)\wedge\nabla_m\beta,
    \end{align}
    that is only the crossed terms in the expansion of $\delta(\alpha\wedge\beta)$ with respect to $\nabla$ remain.
\end{remark}
\begin{remark}
    Equation~\eqref{eq:delta_Schouten} can be viewed as the commutator between $\delta$ and $j^\alpha$ and is registered as such in Eq.~\eqref{eq:[Delta,j^a]}.
    Also, it entails a nice identity for the Laplace--de Rham operator, $\square = -(d\delta + \delta d)$, acting on a product of two forms
    \begin{equation}
        \label{eq:squareAB}
        \square(\alpha\wedge\beta)
        = (\square\alpha)\wedge\beta
        + \alpha\wedge(\square\beta)
        - \llbra\alpha,d\beta\rrket
        + (-1)^a\llbra d\alpha,\beta\rrket
        - (-1)^a d\llbra\alpha,\beta\rrket,
    \end{equation}
    which can also be viewed as a commutator, see Eq.~\eqref{eq:[square,j^a]}.
    In the case that one of the two forms in the product is a function or a $1$-form Eq.~\eqref{eq:squareAB} simplifies to the cases as follows:
    \begin{alignat}{2}
        \square(\phi\psi)
        &=\psi(\square\phi)
        +\phi(\square\psi)
        + 2(d\phi,d\psi),
        &&\phi,\psi\in\Omega^0,\\
        \square(\phi\lambda)
        &= \phi(\square\lambda)
        + \Lie_{\sharp d\phi}\lambda
        + * ^{-1}\Lie_{\sharp d\phi}*\lambda
        &&\\
        &=(\square\phi)\lambda
        + \Lie_{\sharp d\phi}\lambda
        - (\delta\lambda)d\phi
        + \delta(\lambda\wedge d\phi),
        &\quad&\phi\in\Omega^0,\ \lambda\in\Omega^1,\\
        \square(\phi\beta)
        &= (\square\phi)\beta
        +\phi(\square\beta)
        + \Lie_{\sharp d\phi}\beta
        + \flat\Lie_{\sharp d\phi}\sharp\beta
        &&\\
        &=\phi(\square\beta)
        + \Lie_{\sharp d\phi}\beta
        + *^{-1}\Lie_{\sharp d\phi}*\beta,
        &&\phi\in\Omega^0,\ \beta\in\Omega^b,\\
        \square(\lambda\wedge\tau)
        &= (\square\lambda)\wedge\tau
        + \lambda\wedge(\square\tau)
        -\flat\Lie_{\sharp\lambda}\sharp d\tau
        + \flat\Lie_{\sharp\tau}\sharp d\lambda
        + d\flat\Lie_{\sharp\lambda}\sharp\tau
        &&\\
        &= -(\delta d\lambda)\wedge\tau
        + \lambda\wedge(\square\tau)
        + (\delta\tau)\lambda &&\nonumber\\ &\qquad
        - *^{-1}\Lie_{\sharp\lambda}*d\tau
        + *^{-1}\Lie_{\sharp\tau}*d\lambda
        + d*^{-1}\Lie_{\sharp\lambda}*\tau,
        &&\lambda,\tau\in\Omega^1.
    \end{alignat}
\end{remark}

\section{A generalized interior product and its interplay with the Schouten--Nijenhuis bracket}
\label{sec:Produit_Generalise}

\subsection{An interior product naturally generalized with respect to the Hodge operator}

We want to extend the interior and exterior products $i^\lambda$ and $j^\lambda$ from $\lambda\in\Omega^1$ to any form $\alpha\in\Omega^a$.
The extension of $j$ is very natural: we note
\begin{equation}
    j^\alpha\beta=\alpha\wedge\beta,
\end{equation}
and we ask for $i^\alpha$ to correspond to $j^\alpha$ as in Eq.~\eqref{eq:springboard1} through
\begin{equation}
    i^\alpha\beta=\pm\ast^{-1}j^\alpha\ast\beta,
\end{equation}
tweaking its sign such that it coincides with the $a=1$ case and that $i^\phi\alpha=\phi\alpha$ for $\phi\in\Omega^0$.
These constraints lead necessarily to the definition below. 
Various generalized products have been introduced\cite{Koszul:1985,Marle97,Coll:2003ym} and can be introduced, here we advocate for the ``naturalness'' of the present definition with respect to the Hodge operator.

\begin{definition}\label{def:Prod_Gene}
    Let $\alpha\in\Omega^a$ and $\beta\in\Omega^b$ be two forms, we define the generalized interior product as
    \begin{equation}\label{eq:Prod_Gene}
        i^\alpha\beta =
            (-1)^{a(b+1)}*^{-1}j^\alpha*\beta
            \in\Omega^{b-a}.
    \end{equation}
    As a consequence we have $i^\alpha\beta =0$ whenever $a>b$ and $i^\alpha\beta =(\alpha,\beta)$ whenever $a=b$.
\end{definition}

For $\alpha=\flat v\in\Omega^1$ a $1$-form it agrees with the usual interior product (Eq.~\eqref{eq:springboard1}) and, through direct computations, it also fulfills:
\begin{align}
    &i^\phi\alpha = \phi\alpha,\label{eq:Id-I^alpha-J1}\\
    &i^\alpha\omega = *\alpha,\\
    &*i^\alpha\beta = (-1)^{a(b+1)}j^\alpha*\beta,\label{eq:Id-I^alpha-J4}\\
    &i^\alpha*\beta = (-1)^{ab}*j^\alpha\beta,\label{eq:Id-I^alpha-J3}\\
    &i^\alpha i^\beta\gamma
        = i^{\beta\wedge\alpha}\gamma
        = (-1)^{ab}i^{\alpha\wedge\beta}\gamma
        = (-1)^{ab}i^\beta i^\alpha\gamma,\label{eq:iAiB}\\
    &i^\alpha*\beta = (-1)^{ab}i^\beta*\alpha,\\
    &i^{*\alpha}\beta = (-1)^{a+b+ab + n} i^{*\beta}\alpha,\\
    &i^{*\alpha}*\beta = \sgn(g)(-1)^{n(a+b) + a(b+1)} i^\beta\alpha,\label{eq:Id-I^*alpha*beta}
\end{align}
for $\alpha\in\Omega^a$, $\beta\in\Omega^b$ and $\gamma\in\Omega^c$ forms, $\phi$ a function and $\omega$ the volume-form.

\begin{property}
    Let $\alpha\in\Omega^a$ and $\beta\in\Omega^b$ be two forms such that $a\leq b$,
    then the interior product is explicitly given by
    \begin{equation}
        i^\alpha\beta = \frac{1}{a!(b-a)!}\alpha_{m_1...m_a}\beta_{n_1...n_b}
            \tilde{g}(e^{m_1},e^{n_1})\cdots\tilde{g}(e^{m_a},e^{n_a})
            e^{n_{a+1}}\wedge\cdots\wedge e^{n_b},
    \end{equation}
    with $\tilde{g}(e^a,e^b) = g(\sharp e^a, \sharp e^b) = \langle e^a,\sharp e^b\rangle$.
    In an orthonormal basis it reads
    \begin{equation}
        i^\alpha\beta = \frac{1}{a!(b-a)!}\alpha_{m_1...m_a}\beta_{n_1...n_b}
            \eta^{m_1n_1}\cdots\eta^{m_an_a}
            e^{n_{a+1}}\wedge\cdots\wedge e^{n_b}.
    \end{equation}
\end{property}
\begin{proof}
    Direct computation using the fact that $e^{m_1}\wedge\cdots\wedge e^{m_a}\wedge\cdot = j^{e^{m_1}}\cdots j^{e^{m_a}}\cdot\equiv j^{m_1}\cdots j^{m_a}\cdot$ and commuting $*$ and $j$ using Eq.~\eqref{eq:*j_vA}, in order to get
    \begin{equation}
        i^\alpha\beta = \frac{1}{a!}\alpha_{m_1\cdots m_a}i^{m_a}\cdots i^{m_1}\beta.
    \end{equation}
    Then, by writing $\beta = \frac{1}{b!}\beta_{n_1\cdots n_b} j^{n_1}\cdots j^{n_b}1$ and 
    using Eq.~\eqref{eq:i_ujv} to anti-commute the $i$s and the $j$s, thus generating the $g(\sharp e^{m_r},\sharp e^{n_s})= \tilde{g}(e^{m_r}, e^{n_s})$ terms, one obtains the end result.
\end{proof}

Unfortunately, for generic forms $\alpha\in\Omega^a$ and $\beta\in\Omega^b$ there is no remarkable identity  for $i^\alpha j^\beta$, because of the loss of a Leibniz identity.
However, for $\lambda\in\Omega^1$ a $1$-form, there remains the two special cases
\begin{align}
    &[i^\alpha,j^\lambda]_a\beta
    = i^{i^\lambda\alpha}\beta,\label{eq:iiLAB}\\
    &[i^\lambda,j^\alpha]_a \beta
    = j^{i^\lambda\alpha}\beta.
\end{align}

\subsection{Combining the generalized product and the SN bracket}

The generalized interior product can be used through the first few equations it fulfills, Eqs.~\eqref{eq:Id-I^alpha-J1}--\eqref{eq:Id-I^*alpha*beta}.
It can also interact with the Schouten--Nijenhuis bracket, see Eqs.~\eqref{eq:i[A,B]C} and \eqref{eq:iL[A,B]}.
More importantly here, while dealing with computations involving differential operators on $p$-forms one is led to use (anti-)commutations relations between such operators.
Then both the generalized interior product and the SN bracket necessarily show up in combination.
Equations~\eqref{eq:[d,jA]B}--\eqref{eq:[iA,box]B}, and following, deal with such computations
and are the salient results of this study, furthering the toolbox of identities used to compute with $p$-forms.
From these identities one can obtain additional ones as special cases (see e.g. Eqs.~\eqref{eq:[Lv,delta]_1} and \eqref{eq:[Lv,delta]_2}).

Now, the interior product can also either act on a bracket or take a bracket as its argument:
\begin{align}
    &i^{\llbra\alpha,\beta\rrket}\gamma
    = (-1)^a i^{\alpha\wedge\beta}d\gamma
    + (-1)^b d i^{\alpha\wedge\beta}\gamma
    + (-1)^{(a-1)(b-1)}i^\alpha d i^\beta\gamma
    - i^\beta d i^\alpha \gamma,\label{eq:i[A,B]C}\\
    &i^\lambda\llbra\alpha,\beta\rrket
    = \llbra i^\lambda\alpha,\beta\rrket
    -(-1)^a\llbra\alpha,i^{\lambda}\beta\rrket
    + (-1)^a[i^{d\lambda}\alpha\wedge\beta
    -(i^{d\lambda}\alpha)\wedge\beta
    - \alpha\wedge(i^{d\lambda}\beta)],\label{eq:iL[A,B]}
\end{align}
with $\alpha\in\Omega^a$, $\beta\in\Omega^b$ and $\gamma\in\Omega^c$ and the latter equation holds only for $\lambda\in\Omega^1$ a $1$-form.
For $\lambda = d\phi\in\Omega^1$ an exact $1$-form it simplifies strikingly to
\begin{equation}
    i^{d\phi}\llbra\alpha,\beta\rrket
    = \llbra i^{d\phi}\alpha,\beta\rrket
    -(-1)^a\llbra\alpha,i^{d\phi}\beta\rrket,
\end{equation}
and can be recovered using the graded Jacobi identity Eqs.~\eqref{eq:Jacobi_gradue} and \eqref{eq:Bracket_Scalaire_FormeGene}.
In a similar vein setting $\alpha=\phi\in\Omega^0$ a function in Eq.~\eqref{eq:i[A,B]C} simplifies it to
\begin{equation}
    i^{\llbra\phi,\beta\rrket}\gamma
    = -i^{i^{d\phi}\beta}\gamma
    = -i^\beta (d\phi)\wedge\gamma 
    +(-1)^b(d\phi)\wedge i^\beta\gamma
    = -[i^\beta,j^{d\phi}]_b\gamma,
\end{equation}
and can be recovered, then, directly from Eqs.~\eqref{eq:iiLAB} and \eqref{eq:Bracket_Scalaire_FormeGene}.

Of interest for practical computations are the various commutators involving differential operators with $j^\alpha$ and $i^\alpha$:
\begin{align}
    &[d,j^\alpha]_a\beta = j^{d\alpha}\beta,\label{eq:[d,jA]B}\\
    &[\delta,j^\alpha]_a\beta
    = j^{\delta\alpha}\beta
    + (-1)^a\llbra\alpha,\beta\rrket,\label{eq:[Delta,j^a]}\\
    &[\Lie_v,j^\alpha]\beta = j^{\Lie_v\alpha}\beta,\label{eq:[LieV,jA]B}\\
    &[\square, j^\alpha]\beta
    = j^{\square\alpha}\beta
    - \llbra\alpha,d\beta\rrket
    + (-1)^a\llbra d\alpha,\beta\rrket
    - (-1)^a d\llbra\alpha,\beta\rrket,\label{eq:[square,j^a]}\\
    &[i^\alpha, d]_a\beta
    = -i^{\delta\alpha}\beta
    + (-1)^{b(a+1)}*^{-1}\llbra\alpha,*\beta\rrket,\label{eq:[iA,d]B}\\
    &[i^\alpha,\delta]_a \beta = i^{d\alpha}\beta,\\
    &[i^\alpha,\Lie_v]\beta 
        = -i^{\llbra\flat v,\alpha\rrket}\beta
        = -i^{\flat\Lie_{v}\sharp\alpha}\beta
        = -i^{*^{-1}\Lie_v*\alpha}\beta - \mathrm{div}(v)i^\alpha\beta,
        \label{eq:[iA,LieV]B}\\
    &[i^\alpha,\square]\beta
    = -i^{\square\alpha}\beta
    -(-1)^{ab} *^{-1}\llbra d\alpha,*\beta\rrket
    +(-1)^{ab} *^{-1}d\llbra\alpha,*\beta\rrket
    +(-1)^{ab+a} *^{-1}\llbra\alpha,d*\beta\rrket,\label{eq:[iA,box]B}
\end{align}
in which the Schouten--Nijenhuis bracket appears naturally in the right-hand side.
Equation~\eqref{eq:[iA,LieV]B} is obtained from Eq.~\eqref{eq:i[A,B]C} by specializing $\lambda = \flat v\in\Omega^1$ a $1$-form thus leading to 
\begin{equation}
    i^{\llbra\lambda,\alpha\rrket}\beta
    = -i^{\lambda\wedge\alpha}d\beta
    + (-1)^ad i^{\lambda\wedge\alpha}\beta
    + i^\lambda d i^\alpha\beta
    - i^\alpha d i^\lambda\beta,
\end{equation}
then using Cartan formula in conjunction with Eq.~\eqref{eq:iAiB} leads directly to the commutator of the Lie derivative with $i^\alpha$.
Similarly, setting $\alpha = \flat v\in\Omega^1$ a $1$-form in
 Eq.~\eqref{eq:[iA,box]B} produces another notable identity between the Lie derivative and the codifferential:
\begin{align}
    [\Lie_v,\delta]\beta
    &= [\square, i_v]\beta
    + [d, i^{d\flat v}]\beta\nonumber\\
    &= [\square, i_v]\beta
    + i_{\sharp\delta d\flat v}\beta
    +(-1)^b *^{-1}\llbra d\flat v,*\beta\rrket,\label{eq:[Lv,delta]}
\end{align}
with the very special case for $\lambda = \flat v$ with $v$ a Killing vector
\begin{equation}\label{eq:iVboxB}
    [i^\lambda,\square]\beta = (d i^{d\lambda} - i^{d\lambda}d)\beta.
\end{equation}

\begin{remark}
    Some miscellaneous identities can be obtained, such as
    \begin{equation}
        \llbra\phi_a,\cdots\llbra\phi_1,\alpha\rrket\cdots\rrket
        =(-1)^ai^{d\phi_a}\cdots i^{d\phi_1}\alpha
        =(-1)^a i^{d\phi_1\wedge\cdots\wedge d\phi_a}\alpha
        =(-1)^a (d\phi_1\wedge\cdots\wedge d\phi_a,\alpha),
    \end{equation}
    with $\phi_1$, ..., $\phi_a$ functions and $\alpha\in\Omega^a$ a form of degree $a$.
    Also, for $\lambda_1$ and $\lambda_2$ two $1$-forms, one easily gets from Eq.~\eqref{eq:Jacobi_gradue}
    that
    \begin{equation}
        \llbra\lambda_1,\llbra\lambda_2,\alpha\rrket\rrket
        - \llbra\lambda_2,\llbra\lambda_1,\alpha\rrket\rrket
        = \llbra\lambda_3,\alpha\rrket,
    \end{equation}
    with $\lambda_3 = \flat[\sharp\lambda_1,\sharp\lambda_2]$.
\end{remark}
Finally, it can be interesting to introduce Tulczyjew\cite{Tulczyjew:1974}, or Michor\cite{Michor1987}, differential operator
\begin{equation}\label{eq:ML_def}
    \ML^\alpha = [i^\alpha,d]_a = (i^\alpha d -(-1)^adi^\alpha),
\end{equation}
cf. Eq.~\eqref{eq:[iA,d]B} relating $\Theta$ to the Schouten--Nijenhuis bracket, fulfilling
\begin{alignat}{2}
    &\ML^\phi\alpha = -(d\phi)\wedge\alpha = -j^{d\phi}\alpha, 
        &&\phi\in\Omega^0,\ \alpha\in\Omega^a,\label{eq:ML_eq1}\\
    &\ML^\lambda\alpha = \Lie_{\sharp\lambda}\alpha,
    &&\lambda\in\Omega^1,\ \alpha\in\Omega^a,\\
    &\ML^{\alpha\wedge\beta}
    = i^\beta\ML^\alpha + (-1)^a\ML^\beta i^\alpha
    = (-1)^{ab}[i^\alpha\ML^\beta + (-1)^b\ML^\alpha i^\beta]
        &&\\
    &\phantom{\ML^{\alpha\wedge\beta}\,}
        =[i^\beta,\ML^\alpha]_{a+1}
        = (-1)^{ab}[i^\alpha,\ML^\beta]_{b+1},
        &&\alpha\in\Omega^a,\ \beta\in\Omega^b,\\
    &[\ML^\alpha,d]_{a+1} = 0,
        &&\alpha\in\Omega^a,\label{eq:ML_eq4}\\
    &i^{\llbra\alpha,\beta\rrket}\gamma
        = (-1)^a i^\beta\ML^\alpha\gamma
        + (-1)^{(a-1)(b-1)}\ML^\alpha i^\beta\gamma
        &&\label{eq:ML_eq5}\\
    &\phantom{i^{\llbra\alpha,\beta\rrket}\gamma\,}
        =(-1)^a[i^\beta,\Theta^\alpha]_{b(a+1)}\gamma
        = -[\ML^\beta,i^\alpha]_{a(b+1)}\gamma,
        &&\alpha\in\Omega^a,\ \beta\in\Omega^b,\ \gamma\in\Omega^c,\\
    &\ML^{\llbra\alpha,\beta\rrket}
    = (-1)^{(a-1)(b-1)}\ML^\alpha\ML^\beta
    - \ML^\beta\ML^\alpha
    = -[\ML^\beta,\ML^\alpha]_{(a-1)(b-1)},\ 
        &&\alpha\in\Omega^a,\ \beta\in\Omega^b.\label{eq:ML_eq6}
\end{alignat}
Out of these identities one can notice the special cases:
\begin{equation}
    \ML^{\alpha\wedge\beta}\gamma 
    = \begin{cases}
        \ [i^\beta,\ML^\alpha]\gamma = -i^{\llbra\alpha,\beta\rrket}\gamma,\quad
        & \text{$a$ odd},\\[.5em]
        \ [i^\beta,\ML^\alpha]_+\gamma = i^{\llbra\alpha,\beta\rrket}\gamma,
        & \text{$a$ even, $b$ odd},\\[.5em]
        \ i^{\llbra\alpha,\beta\rrket}\gamma + 2\ML^\alpha i^\beta\gamma,
        & \text{$a$ even, $b$ even.}
    \end{cases}
\end{equation}
It also enables to recast Eq.~\eqref{eq:[Lv,delta]} as
\begin{equation}\label{eq:[Lv,delta]_1}
    [\Lie_v,\delta]\beta 
    = [\square,i_v]\beta
    - \ML^{d\flat v}\beta,
\end{equation}
and in the special case of $v = \sharp\lambda$ a Killing vector it simplifies to
\begin{equation}\label{eq:[Lv,delta]_2}
    [\square,i^\lambda]\beta
    = \ML^{d\lambda}\beta.
\end{equation}

\appendix

\section{Overview of needed formulas of differential geometry}
\label{app:Overview_Geo_Diff}

Let $(M,g)$ be a smooth pseudo-Riemannian manifold of dimension $n$,
$TM$ and $T^*M$ being its tangent and cotangent fiber bundles, respectively.
For $p$ an integer such that $1\leq p\leq n$ then $\Omega^p(M)\equiv \Omega^p = \smash{\bigwedge^p T^*M}$ is the space of $p$-forms, smooth functions on $M$ are identified with zero-forms: $\Omega^0(M) = C^{\infty}(M,\mathbb{R})$.

The natural pairing between a $1$-form $\lambda$ and a vector $v$ is written $\langle\lambda,v\rangle$, the musical applications $\flat$ and $\sharp$ relate this pairing to the metric as $\langle\lambda,v\rangle = g(\sharp\lambda,v)$ and 
$g(u,v) = \langle\flat u,v\rangle.$
Of prime importance here are the ``creator operator'' $j_v$, such that $j_v\alpha = (\flat v)\wedge\alpha$ with $v\in TM$ and $\alpha\in\Omega^a$, and the interior product $i_v$ (or insertion operator $i_v$) fulfilling an antisymmetry relation $i_ui_v = -i_vi_u$, a Leibniz identity
\begin{equation}\label{eq:Leibniz_iv}
    i_v\alpha\wedge\beta = (i_v\alpha)\wedge\beta + (-1)^a\alpha\wedge(i_v\beta),\quad
    \alpha\in\Omega^a,\ \beta\in\Omega^b,
\end{equation}
and interacting with $j_v$ as
\begin{equation}\label{eq:i_ujv}
    i_u j_v + j_vi_u = g(u,v),\quad u,v\in TM.
\end{equation}
We recall that the interior product appears in the Lie derivative through Cartan formula: $\Lie_v = i_v d + di_v$.

On an oriented manifold with the Hodge operator $*$, fulfilling $*1=\omega$,  $*\omega = \mathrm{sgn}(g)$ with $\omega$ the volume form, and $**\alpha = \sgn(g)(-1)^{a(n+1)}\alpha$ for $\alpha\in\Omega^a$,
the product between $p$-forms is defined as
\begin{equation}
    (\alpha,\beta) = *^{-1}(\alpha\wedge*\beta),\quad \alpha,\beta\in\Omega^p,
\end{equation}
and the codifferential as $\delta\alpha = (-1)^a*^{-1}d*\alpha$.
For $\alpha\in\Omega^a$ a $a$-form, $*$ interacts with $i_v$ and $j_v$ as
\begin{align}
    &*i_v\alpha = -(-1)^aj_v*\alpha,\label{eq:*i_vA}\\
    &*j_v\alpha = (-1)^ai_v*\alpha,\label{eq:*j_vA}
\end{align}
see, say, App.~A of Ref.~\onlinecite{Huguet:2022rxi} for additional identities.
The generalized interior product stems from Eq.~\eqref{eq:*i_vA}.

\section{Proofs and sketches of a few long computations}
\label{app:Long_Computations}

\subsection{Proof of Eq.~\eqref{eq:L_gtilde}}

\begin{proof}
    First, notice that both $\llbra\lambda,\beta\rrket -\Lie_{\sharp\lambda}\beta$ and
    $(\Lie_{\sharp\lambda}\tilde g)^{mn}j_m i_n\beta$ vanish on functions (i.e. $0$-forms) and agree on $1$-forms.
    Indeed, on a $1$-form $\tau\in\Omega^1$ one has $\llbra\lambda,\tau\rrket - \Lie_{\sharp\lambda}\tau = \flat\Lie_{\sharp\lambda}\sharp\tau - \Lie_{\sharp\lambda}\tau$ but
    \begin{align}
        \flat\Lie_{\sharp\lambda}\sharp\tau
        &= \flat \Lie_{\sharp\lambda}C(\tilde{g},\tau)
        = \flat C(\Lie_{\sharp\lambda}\tilde{g},\tau)
        + \flat C(\tilde{g},\Lie_{\sharp\lambda}\tau)\nonumber\\
        &= \flat C(\Lie_{\sharp\lambda}\tilde{g},\tau)
        + \Lie_{\sharp\lambda}\tau=
        (\Lie_{\sharp\lambda}\tilde g)^{mn}j_m i_n\tau
        + \Lie_{\sharp\lambda}\tau,
    \end{align}
    where $C$ is the contraction of a symmetric 2-vector with a 1-form.
   
    Furthermore, we note that $\llbra\lambda,\beta\rrket -\Lie_{\sharp\lambda}\beta$ and
    $(\Lie_{\sharp\lambda}\tilde g)^{mn}j_m i_n\beta$ are both derivatives on $p$-forms fulfilling the same Leibniz identity (derivation of degree 0 on the graded algebra $\Omega$).
    Indeed, the same Leibniz identity holds for the three terms above: for $\llbra\lambda,\cdot\rrket$ thanks to Eq.~\eqref{eq:Schouten_triple},
    for $\Lie_{\sharp\lambda}$ it is the usual one and for $j_mi_n$ thanks to Eq.~\eqref{eq:Leibniz_iv} and to the fact that $j_m\alpha\wedge\beta = (-1)^a\alpha\wedge j_m\beta$ for $\alpha\in\Omega^a$ and $\beta\in\Omega^b$.
    Finally, these two derivations of degree zero are equal since they agree on 0-forms and 1-forms.
\end{proof}

\subsection{Proof of Prop.~\ref{prop:Crochet_explicite}}

\begin{proof}
    Expanding first $\alpha$, using Eq.~\eqref{eq:AB-C_gauche}, brings about
    \begin{equation}\label{eq:Dvlpt_Schouten_Gauche}
        \llbra\alpha,\beta\rrket
        = \frac{1}{a!}\left\{
        e^{m_1}\wedge\cdots\wedge e^{m_a}\wedge
        \llbra\alpha_{m_1\cdots m_a},\beta\rrket
        + a \alpha_{m_1\cdots m_a}
        e^{m_1}\wedge\cdots\wedge e^{m_{a-1}}\wedge
        \llbra e^{m_a},\beta\rrket
        \right\},
    \end{equation}
    using the skew-symmetry of the coefficient.
    Then, permuting the arguments in the brackets and using again this formula, gives rise to
    \begin{multline}
        \llbra\alpha,\beta\rrket
        = \frac{1}{a!b!}\Bigl\{
        -(-1)^b b\llbra\alpha_{m_1\cdots m_a}, e^{n_b}\rrket
        \beta_{n_1\cdots n_b}
        e^{m_1}\wedge\cdots\wedge e^{m_a}\wedge
        e^{n_1}\wedge\cdots\wedge e^{n_{b-1}}\\
        - a \alpha_{m_1\cdots m_a} \llbra\beta_{n_1\cdots n_b}, e^{m_a}\rrket
        e^{m_1}\wedge\cdots\wedge e^{m_{a-1}}\wedge
        e^{n_1}\wedge\cdots\wedge e^{n_b}\\
        + ab \alpha_{m_1\cdots m_a}\beta_{n_1\cdots n_b}
        e^{m_1}\wedge\cdots\wedge e^{m_{a-1}}\wedge
        e^{n_1}\wedge\cdots\wedge e^{n_{b-1}}\wedge
        \llbra e^{m_a}, e^{n_b}\rrket
        \Bigr\}.
    \end{multline}
    Shuffling the indices and relabeling them in a straightforward manner, with $c=a+b-1$, simplifies to
    \begin{multline}
        \llbra\alpha,\beta\rrket
        = \frac{1}{a!b!}\Bigl\{
        b\llbra\alpha_{k_1\cdots k_{a}}, e^{r}\rrket
        \beta_{k_{a+1}\cdots k_c r}
        - a \alpha_{k_1\cdots k_{a-1} r} \llbra\beta_{k_a\cdots k_c}, 
        e^{r}\rrket\\
        + ab \alpha_{k_1\cdots k_{a-1}r}\beta_{k_a\cdots k_{c-1} s}
        \left(\llbra e^{r}, e^{s}\rrket\right)_{k_c}
        \Bigr\}e^{k_1}\wedge\cdots\wedge e^{k_c}.
    \end{multline}
    Using, in addition, Eq.~\eqref{eq:Bracket_Scalaire_1forme}
    to express $\llbra\alpha_{k_1\cdots k_{a}}, e^{r}\rrket$ and $\llbra\beta_{k_a\cdots k_c}, 
        e^{r}\rrket$
    and that
    $\llbra e^r,e^s\rrket 
        = \flat\Lie_{\sharp e^r}\sharp e^s
       = \tilde c^{rs}_ke^k$
    gives the final formula.
\end{proof}

\subsection{Proof of Prop.~\ref{thm:delta_Schouten}}

\begin{proof}
    First let $a=0$, that is $\alpha=\phi\in\Omega^0$ be a function and $\beta\in\Omega^b$ a $b$-form, 
    then
    \begin{equation}\label{eq:Dab_0}
        \delta(\phi\beta)
        = \delta(\phi\wedge\beta)
        = \phi(\delta\beta)
        + \llbra\phi,\beta\rrket,
    \end{equation}
    which follows from the definition of the codifferential $\delta\beta = (-1)^b *^{-1}d*\beta$,
    and that, on the one hand, one has
    $\llbra\phi,\beta\rrket = - i^{d\phi}\beta$ (cf. Eq.~\eqref{eq:Bracket_Scalaire_FormeGene})
    and, on the other hand, one has $i^{d\phi}\beta = \phi\delta\beta - \delta(\phi\beta)$, combining both equations proves the formula for $a=0$.
    
    Now let $a=1$, that is $\alpha$ being a $1$-form, and $\beta\in\Omega^b$ a $b$-form, we then have
    \begin{equation}
        \delta(\alpha\wedge\beta)
        = -\alpha\wedge\delta\beta
        - *^{-1}\Lie_{\sharp\alpha}*\beta,
    \end{equation}
    by commuting $j^\alpha$ with $*$ in $\delta$ with Eq.~\eqref{eq:*j_vA} and using Cartan formula to recast $d i^\alpha$ as $\Lie_{\sharp\alpha} - i^\alpha d$ and finally commuting the remaining $i^\alpha$ with $*^{-1}$ with Eq.~\eqref{eq:*i_vA}.
    Then owing to Eq.~\eqref{eq:*L*B} this reads as
    \begin{equation}\label{eq:Dab_1}
        \delta(\alpha\wedge\beta)
        = -\alpha\wedge\delta\beta
        - (\llbra\alpha,\beta\rrket
        - (\delta\alpha)\beta)
        = (\delta\alpha)\beta - \alpha\wedge\delta\beta - \llbra\alpha,\beta\rrket,
    \end{equation}
    thus proving the formula for $a=1$.
    
    The rest of the proof relies on an induction on $a$.
    Indeed, assume that the formula holds at an arbitrary $a$, which is already the case for $a=0$ and $a=1$ (cf. Eqs.~\eqref{eq:Dab_0} and \eqref{eq:Dab_1}), and consider $\alpha\in\Omega^{a+1}$ a $(a+1)$-form.
    Let $\beta\in\Omega^b$ be a $b$-form, then
    \begin{align*}
        \delta(\alpha\wedge\beta)
        &= \delta(\alpha_{m_1\cdots m_{a+1}} e^{m_1}\wedge\cdots\wedge e^{m_a}\wedge e^{m_{a+1}}\wedge\beta)\\
        &= (\delta \alpha_{m_1\cdots m_{a+1}} e^{m_1}\wedge\cdots\wedge 
        e^{m_a})\wedge e^{m_{a+1}}\wedge\beta \\&\qquad
        +(-1)^a \alpha_{m_1\cdots m_{a+1}}e^{m_1}\wedge\cdots\wedge e^{m_a}\wedge\delta(e^{m_{a+1}}\wedge\beta)\\ &\qquad
        +(-1)^a \llbra \alpha_{m_1\cdots m_{a+1}} e^{m_1}\wedge\cdots\wedge e^{m_a}, e^{m_{a+1}}\wedge\beta\rrket\\
        &= (\delta \alpha_{m_1\cdots m_{a+1}} e^{m_1}\wedge\cdots\wedge 
        e^{m_a})\wedge e^{m_{a+1}}\wedge\beta \\&\qquad
        +(-1)^a \alpha_{m_1\cdots m_{a+1}}e^{m_1}\wedge\cdots\wedge e^{m_a}\wedge\delta(e^{m_{a+1}})\wedge\beta\\ &\qquad
        +(-1)^{a+1} \alpha_{m_1\cdots m_{a+1}}e^{m_1}\wedge\cdots\wedge 
        e^{m_a}\wedge e^{m_{a+1}}\wedge\delta\beta \\&\qquad
        +(-1)^{a+1} \alpha_{m_1\cdots m_{a+1}}e^{m_1}\wedge\cdots\wedge e^{m_a}\wedge\llbra e^{m_{a+1}},\beta\rrket\\ &\qquad
        +(-1)^a \llbra \alpha_{m_1\cdots m_{a+1}} e^{m_1}\wedge\cdots\wedge e^{m_a}, e^{m_{a+1}}\wedge\beta\rrket\\
        &= (\delta \alpha_{m_1\cdots m_{a+1}} e^{m_1}\wedge\cdots\wedge 
        e^{m_a}\wedge e^{m_{a+1}})\wedge\beta \\&\qquad
        + (-1)^{a+1}\llbra\alpha_{m_1\cdots m_{a+1}} e^{m_1}\wedge\cdots\wedge e^{m_a}, e^{m_{a+1}}\rrket\wedge\beta\\ &\qquad
        +(-1)^{a+1} \alpha\wedge\delta\beta  \\&\qquad
        +(-1)^{a+1} \alpha_{m_1\cdots m_{a+1}}e^{m_1}\wedge\cdots\wedge e^{m_a}\wedge\llbra e^{m_{a+1}},\beta\rrket\\ &\qquad
        +(-1)^a \llbra \alpha_{m_1\cdots m_{a+1}} e^{m_1}\wedge\cdots\wedge e^{m_a}, e^{m_{a+1}}\wedge\beta\rrket\\
        &= (\delta \alpha)\wedge\beta 
        +(-1)^{a+1} \alpha\wedge\delta\beta \\&\qquad
        + (-1)^{a+1}\llbra\alpha_{m_1\cdots m_{a+1}} e^{m_1}\wedge\cdots\wedge e^{m_a}, e^{m_{a+1}}\rrket\wedge\beta\\ &\qquad
        +(-1)^{a+1} \alpha_{m_1\cdots m_{a+1}}e^{m_1}\wedge\cdots\wedge e^{m_a}\wedge\llbra e^{m_{a+1}},\beta\rrket\\ &\qquad
        +(-1)^a \llbra \alpha_{m_1\cdots m_{a+1}} e^{m_1}\wedge\cdots\wedge e^{m_a}, e^{m_{a+1}}\wedge\beta\rrket\\
        &= (\delta \alpha)\wedge\beta
        +(-1)^{a+1} \alpha\wedge\delta\beta \\&\qquad
        +(-1)^{a+1} \alpha_{m_1\cdots m_{a+1}}e^{m_1}\wedge\cdots\wedge e^{m_a}\wedge\llbra e^{m_{a+1}},\beta\rrket\\ &\qquad
        + e^{m_{a+1}}\wedge\llbra\alpha_{m_1\cdots m_{a+1}} e^{m_1}\wedge\cdots\wedge e^{m_a}, \beta\rrket\\
        &= (\delta \alpha)\wedge\beta
        +(-1)^{a+1} \alpha\wedge\delta\beta
        +(-1)^{a+1} \llbra\alpha,\beta\rrket,
    \end{align*}
    in which the combinatorial factor $1/(a+1)!$ has been incorporated in $\alpha_{m_1\cdots m_{a+1}}$ as it plays no role here, then
    the recurrence hypothesis is used in the first to second line,
    Eq.~\eqref{eq:Dab_1} is used on the second term in the second to third line,
    the recurrence hypothesis is again used to recognize $(\delta\alpha)\wedge\beta$ and thus creating a SN bracket in the third to fourth line.
    The fifth line is a rewriting of the fourth putting first the sought for terms and keeping the SN brackets to reshape as additional terms.
    Equation~\eqref{eq:Schouten_triple} is used on the last term of the fifth line which simplifies partially with the first SN bracket. 
    Finally Eq.~\eqref{eq:AB-C_gauche} is used to recognize in the last two terms $(-1)^{a+1}\llbra\alpha,\beta\rrket$.
\end{proof}

\subsection{Derivation of \texorpdfstring{Eq.~\eqref{eq:i[A,B]C}}{Eq.~\ref*{eq:i[A,B]C}}}

In order to establish Eq.~\eqref{eq:i[A,B]C} one relies on the representation of the bracket as it appears in Eq.~\eqref{eq:Representation_Bracket} and one is led to consider the term $i^{\delta(\alpha\wedge\beta)}\gamma$. 
Using Eq.~\eqref{eq:AB-C_gauche} and explicit formulas for $\delta$, it reads
\begin{align}
    i^{\delta(\alpha\wedge\beta)}\gamma
    =& (-1)^{(a+b-1)(c+1)}*^{-1}(\delta(\alpha\wedge\beta))\wedge*\gamma\nonumber\\
    =& (-1)^{(a+b-1)c}*^{-1}\{
    (-1)^{a+b+1}\delta(\alpha\wedge\beta\wedge*\gamma)
        + \alpha\wedge\beta\wedge(\delta*\gamma)
        + \llbra\alpha\wedge\beta,*\gamma\rrket
    \}\nonumber\\
    =& (-1)^{(a+b)c}d*^{-1}\alpha\wedge\beta\wedge*\gamma
    + (-1)^{(a+b)c + 1} * ^{-1}\alpha\wedge\beta\wedge*d\gamma\nonumber\\
    &+ (-1)^{(a+b-1)c} *^{-1}\alpha\wedge*(*^{-1}\llbra\beta,*\gamma\rrket)
    + (-1)^{(a+b-1)c + ab} *^{-1}\beta\wedge*(*^{-1}\llbra\alpha,*\gamma\rrket).
\end{align}
Owing to its definition, Eq.~\eqref{eq:Prod_Gene}, we recognize in the above expression the generalized interior products $i^{\alpha\wedge\beta}$, $i^\alpha$ and $i^\beta$.
The remaining terms: $*^{-1}\llbra\beta,*\gamma\rrket$ and  $*^{-1}\llbra\alpha,*\gamma\rrket$, are reexpressed using Eq.~\eqref{eq:[iA,d]B} backward.
Bringing all together with the remaining terms of Eq.~\eqref{eq:Representation_Bracket}: $i^{(\delta\alpha)\wedge\beta}$ and $i^{\alpha\wedge(\delta\beta)}$, produces Eq.~\eqref{eq:i[A,B]C}.

\subsection{Derivation of Eqs.~\texorpdfstring{\eqref{eq:[d,jA]B}--\eqref{eq:[iA,box]B}}{\ref*{eq:[d,jA]B}--\ref*{eq:[iA,box]B}}}

First, notice that among the set of Eqs.~\eqref{eq:[d,jA]B}--\eqref{eq:[iA,box]B} a few are known formulas rewritten as commutators, namely:
Eq.~\eqref{eq:[d,jA]B} is Leibniz identity for the exterior derivative,
Eq.~\eqref{eq:[Delta,j^a]} is Eq.~\eqref{eq:delta_Schouten},
Eq.~\eqref{eq:[LieV,jA]B} is Leibniz identity for the Lie derivative,
Eq.~\eqref{eq:[square,j^a]} is Eq.~\eqref{eq:squareAB} which is obtained from Eqs.~\eqref{eq:[d,jA]B} and \eqref{eq:[Delta,j^a]}.
The remaining formulas, Eqs.~\eqref{eq:[iA,d]B}--\eqref{eq:[iA,box]B}, are found from the previous by conjugation with the Hodge operator: $x\beta \to *^{-1}x *\beta$.

\subsection{Derivation of Eqs.~\texorpdfstring{\eqref{eq:ML_eq1}--\eqref{eq:ML_eq6}}{\ref*{eq:ML_eq1}--\ref*{eq:ML_eq6}}}

The first few equations in Eqs.~\eqref{eq:ML_eq1}--\eqref{eq:ML_eq6} are straight consequences of the definition~\eqref{eq:ML_def} of $\Theta^\alpha$ and that of the signed commutator along with Cartan and Leibniz identities, taking into account $d^2 = 0$ and $i^{\alpha\wedge\beta} = i^\beta i^\alpha = (-1)^{ab}i^\alpha i^\beta$ (cf.~Eq.~\eqref{eq:iAiB}).
Equation~\eqref{eq:ML_eq5} relies on Eq.~\eqref{eq:i[A,B]C} and on Eq.~\eqref{eq:iAiB} to recast, say, $i^\beta d$ as $\ML^\beta + (-1)^bdi^\beta$ thus leading to the result and its various expressions.
Establishing Eq.~\eqref{eq:ML_eq6} starts with the definition~\eqref{eq:ML_def} of $\ML^\alpha$ and by using directly Eq.~\eqref{eq:ML_eq5}, then using Eq.~\eqref{eq:ML_eq4} to commute $\ML$ and $d$ yields the result.

\end{document}